\documentclass[english,12pt]{article}
\usepackage{array}
\usepackage{graphicx}
\usepackage{amssymb}
\usepackage{amsmath}
\usepackage{multirow}
\usepackage{prettyref}
\usepackage{color}
\usepackage{babel}
\usepackage{units}
\usepackage[latin1]{inputenc}
\usepackage{amsfonts}
\usepackage{amssymb}
\usepackage{babel}
\usepackage{cite}
\def\@fmsl@sh#1#2#3{\m@th\ooalign{$\hfil#1\mkern#2/\hfil$\crcr$#1#3$}}
 \def\eq#1\en{\begin{equation}#1\end{equation}}
\def\s[#1,#2]{[#1\stackrel{\star}{,}#2]}
\def\sx[#1,#2]{[#1\stackrel{\star_{x}}{,}#2]}

\newcommand{\nc}{\newcommand}
\nc{\beq}{\begin{equation}}
\nc{\eeq}{\end{equation}}
\newcommand{\bea}{\begin{eqnarray}}  
\newcommand{\eea}{\end{eqnarray}}  
\nc{\beqa}{\begin{eqnarray}}
\nc{\eeqa}{\end{eqnarray}}

\def\bc{\begin{center}}
\def\ec{\end{center}}

\def\to{\rightarrow}

\def\gsim{\mathrel{\mathpalette\atversim>}}

\def\bc{\begin{center}}
\def\ec{\end{center}}

\def\gsim{\mathrel{\rlap{\lower4pt\hbox{\hskip1pt$\sim$}}

    \raise1pt\hbox{$>$}}}       

\def\gsim{\mathrel{\rlap{\lower4pt\hbox{\hskip1pt$\sim$}}
    \raise1pt\hbox{$>$}}}       

\begin{document}
\makeatletter
\def\fmslash{\@ifnextchar[{\fmsl@sh}{\fmsl@sh[0mu]}}
\def\fmsl@sh[#1]#2{%
  \mathchoice
    {\@fmsl@sh\displaystyle{#1}{#2}}%
    {\@fmsl@sh\textstyle{#1}{#2}}%
    {\@fmsl@sh\scriptstyle{#1}{#2}}%
    {\@fmsl@sh\scriptscriptstyle{#1}{#2}}}
\def\@fmsl@sh#1#2#3{\m@th\ooalign{$\hfil#1\mkern#2/\hfil$\crcr$#1#3$}}
\makeatother

\thispagestyle{empty}
\begin{titlepage}
\boldmath
\begin{center}
  \Large {\bf Excursion into Quantum Gravity via Inflation}
    \end{center}
\unboldmath
\vspace{0.2cm}
\begin{center}
{  {\large Xavier Calmet}\footnote{x.calmet@sussex.ac.uk} {\large and} {\large Ver\'onica Sanz}\footnote{v.sanz@sussex.ac.uk}}
 \end{center}
\begin{center}
{\sl Physics $\&$ Astronomy, 
University of Sussex,   Falmer, Brighton, BN1 9QH, United Kingdom 
}
\end{center}
\vspace{5cm}
\begin{abstract}
\noindent
The discovery of B-modes, and their effect on the fit to inflationary parameters, opens a window to explore quantum gravity. In this paper we adopt an effective theory approach to study quantum gravity effects in inflation. We apply this approach to chaotic and $\lambda \phi^4$ inflation, and find that BICEP2 constrains these new operators to values which are consistent with the effective theory approach. This result opens the possibility to study quantum gravity in a systematic fashion, including its  effect on Higgs inflation and other Starobisnky-like models. \end{abstract}  
\end{titlepage}



\newpage

\section{ Introduction}
The recent results from the BICEP2 experiment \cite{Ade:2014xna} and their discovery of  B-modes in the cosmic microwave background (CMB) have profound consequences for cosmology and particle physics. The measurement of a 
tensor/scalar ratio $r=0.2^{+0.07}_{-0.05}$, if confirmed by future observations, is one of the most exciting results in physics since the discovery of the black body radiation about 100 years ago. More precisely, the question is truly whether the B-modes observed by BICEP2 are really of cosmological nature or whether they can be explained by a secondary source such as dust. In any case, if correct this result not only supports the hypothesis that our universe went through a period of cosmic inflation in the first few instants of its existence, but it also implies that the energy scale of inflation was very high and close to the Planck scale, around the typical scale for grand unification theories in supersymmetry \cite{Amaldi:1991cn}. Indeed, BICEP2 results imply that inflation took place at an energy scale of $10^{16}$ GeV which according to the Lyth bound \cite{Lyth:1996im} implies that the inflaton field took values much large than the Planck scale.  Inflation is thus sensitive to quantum gravitational physics. We thus for the first time have a chance at probing quantum gravitational physics experimentally by studying the CMB. 

Since we are still very far away from a theory of quantum gravity, we have to rely on an effective theory approach, see e.g. \cite{Calmet:2013hfa,Donoghue:2012zc}. We shall consider the most generic effective theory for a scalar field coupled to gravity and use the recent data from BICEP2 and PLANCK to set limits on the parameters of this effective action for the inflaton $\phi$
\begin{eqnarray}
S=\int d^x \sqrt{-g} \left (\frac{\bar M_P^2}{2} R + f(\phi) F(R,R_{\mu\nu}) + g^{\mu\nu} \partial_\mu \phi \partial^\nu \phi + V_{ren}(\phi)+ \sum_{n=5}^{\infty} c_n \frac{\phi^{n}}{\bar M_P^{n-4}} \right )
\end{eqnarray}
where $\bar M_P$ is the reduced Planck, and $V_{ren}(\phi)$ contains all renormalizable terms up to dimension-four, for example $V_{ren} \supset v^3 \phi+ m^2 \phi^2 + \lambda_3 \phi^3 + \lambda_4 \phi^4$, 
and $c_n$ are Wilson coefficients of the higher-dimensional operators (HDO) to be discussed below.  

The generic term $f(\phi) F(R,R_{\mu\nu})$
stands for non-minimal couplings between the inflaton and the graviton. Via field redefinitions such terms could be shifted in the potential of the scalar field, potentially introducing new scalar degree of freedom as in the case of $R^2$. While these terms can be important for inflation, e.g. in the case of Higgs inflation, we shall ignore them as we wish to consider single field inflationary models. Note that we do not consider derivative terms of the type $(\partial^\mu \phi \partial_\mu \phi \phi^n)$ out of simplicity: we assume that the kinetic term of the graviton is canonically normalized. The validity of using effective theory techniques in inflation has recently been investigated \cite{Collins:2014yua,Agarwal:2013rva}.

We are thus dealing with a potential of the form:
\begin{eqnarray}
V(\phi)=V_{ren} (\phi)+ \sum_{n=5}^{\infty} c_n \frac{\phi^{n}}{\bar M_P^{n-4}}
\end{eqnarray}

  \begin{figure}[t!]
\begin{minipage}{8in}
\hspace*{-0.7in}
\centerline{\includegraphics[height=9cm]{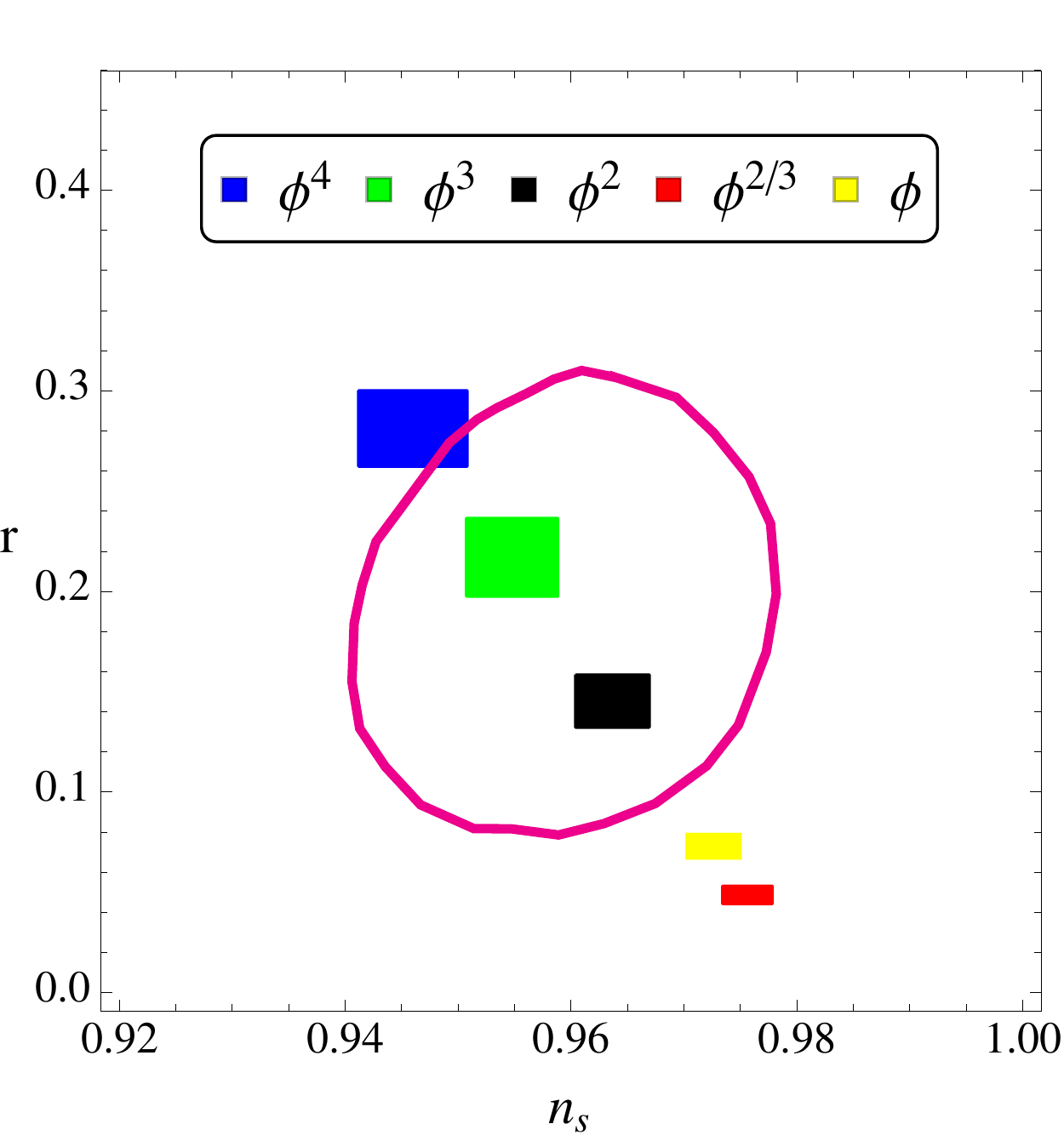}} \label{fignoc}
\hfill
\end{minipage}
\caption{
{\it Predictions for various polynomial forms of $V_{ren}$ with $N \in [50,60]$. The pink circle corresponds to the 95\% CL from BICEP2.
} 
}
\end{figure}

In Fig.~\ref{fignoc} we show the predictions for various polynomial forms of $V_{ren}$. The areas correspond to  $N \in [50,60]$. The pink circle corresponds to the 95\% CL from BICEP2. The models with potentials $\phi^{2-3}$ seem to be favoured, whereas other models with low degree polynomials, or Starobinsky-like seem to be ruled out at the 95\% CL. In this paper we study how the effect of the higher-dimensional operators changes these conclusions.

In inflationary models, one often focusses on one specific term and one sets the remaining Wilson coefficients to zero. However in quantum field theory, with the exception of dimension three and four operators higher dimensional operators will be generated by quantum corrections. The reason why dimension three and four  operators might in principle not be generated in the following. Let's imagine they are indeed generated by quantum gravity. In the limit where $M_P \to \infty$ these operators must go to zero. One thus expects an exponential suppression of such operators by $\exp(-M_P/\mu)$ where  $\mu$ could be some low energy scale \cite{Calmet:2009uz,Holman:1992us}. This reasoning applies to $\phi^3$ and $\phi^4$, if they are not introduced in the action by hand, their Wilson coefficients are expected to be very tiny. On the contrary higher dimensional operators  $c_n \frac{\phi^{n}}{\bar M_P^{n-4}}$ will be generated and one expects that their Wilson coefficients should be of order unity. The precise origin of these higher dimensional operators depends on the underlying theory of quantum gravity. In extra-dimensional theories they could arise from the exchange of super massive KK modes. In a generic theory of quantum gravity, one expects such operators to be generated by virtual and real quantum black holes \cite{Calmet:2011ta}.

This is a well known problem for inflation \cite{Holman:1992us} and it has been argued that these Wilson coefficients should be of order $10^{-3}$ \cite{Kallosh:1995hi,Linde:2007fr}  not to spoil the flatness of the potential. Here we study the implications of the current data on such higher dimensional terms generated by quantum gravity.

\section{The effect of higher-dimensional operators in chaotic inflation}

We consider the potential 
\bea
V(\tilde\phi) = \bar M_P^{4} \left( \tilde m^2 \tilde \phi^2 + c_n \tilde \phi^n \right)
\eea
where the quantities in tilde are normalized to $\bar M_P$, namely
\bea
\tilde \phi = \phi/\bar M_P \,   \ , \,  \tilde m = m/\bar M_P  \ .
\eea
The first term in the potential corresponds to the chaotic inflation model \cite{Linde:1983gd}.

Let us discuss the validity of the expansion in terms of higher-dimensional operators. First, the effect of the first term in the higher-dimensional operator expansion, $c_6$, should be dominant.  Moreover, this higher-dimensional operator should be a correction to the leading renormalizable term on $\tilde m^2$. We then define a parameter $\alpha_m$ as
\bea
c_6 = \alpha_m \tilde m^2 \rightarrow V(\tilde\phi) = \bar M_P^{4} \tilde m^2 \tilde \phi^2  \left( 1+ \alpha_m\tilde \phi^{4} \right) \ ,
\label{aldef}
\eea
which leads to the following condition
\bea
 |\alpha_m|\tilde \phi^{4} < 1
 \label{condexp}
\eea
to perform an expansion.

The presence of the higher-dimensional operator term, or $\alpha_m$, modifies the slow-roll  conditions, 
\bea
\epsilon = \frac{1}{16 \pi} \left(\frac{V'(\tilde\phi)}{V(\tilde \phi)}\right)^2= \frac{1}{4 \pi}\, \frac{1}{\tilde \phi^2} \, \left( \frac{1+3 \, \alpha_m\tilde \phi^{4} }{1+ \alpha_m\tilde \phi^{4}}\right)^2 = \epsilon_{CI}+\frac{\alpha_m \tilde \phi^4}{\pi \tilde \phi^2} + {\cal O}(\alpha_m \tilde \phi^4)^3
\label{eps}
\eea
and  $\epsilon_{CI}=1/(4 \pi \tilde \phi^2)$ is the $\epsilon$ parameter of chaotic inflation, with no higher-dimensional operators. The second slow-roll parameter, $\eta$, which is zero in usual chaotic inflation, receives a contribution from the higher-dimensional operator,
\bea
\eta =\frac{1}{8 \pi} \left( \frac{V''(\tilde\phi)}{V(\tilde\phi)}-\frac{1}{2} \left(\frac{V'(\tilde\phi)}{V(\tilde \phi)}\right)^2 \right) \simeq \frac{5}{2 \pi \tilde \phi^2} \, (\alpha_m\tilde \phi^{4})
\label{eta}
\eea
Similarly, the condition for the end of inflation is then modified to
\bea
\tilde \phi_E^2 = \frac{1}{4 \pi} \, \left(1+  \frac{\alpha_m}{4\pi} \right) \ ,
\eea
and the number of e-foldings,
\bea
N= 2 \sqrt{\pi} \int_{\tilde \phi_E}^{\tilde\phi_I} \frac{1}{\sqrt{\epsilon}} = 2 \pi \tilde\phi_I^2 \left(1-\frac{2\, \alpha_m\tilde \phi_I^{4}}{3}\right)-\frac{1}{2}-\frac{5\,\alpha_m}{48\,\pi^2}
\eea
leading to a value of the value of the field at the beginning of inflation consistent with $N$ e-foldings,
\bea
\tilde \phi_N^2 \simeq  \tilde \phi_{N,CI}^2 + \frac{N^3}{12 \pi^3} \, \alpha_m \simeq \frac{N}{2\pi} \left(1+\frac{N^2 \alpha_m}{6 \pi^2}\right)\,
\eea
where $\tilde \phi_{N,CI}^2=\frac{1+2 N}{4 \pi}$. Although $\alpha_m$ is a small number, the value of the field at the beginning of inflation depends parametrically on $\alpha_m N^2$, where $N$ is a large number, $N\sim 60$. The condition of Eq.~\ref{condexp} then implies
\bea
|\alpha_m| \tilde\phi_N^4 \simeq \frac{N^2 |\alpha_m|}{4 \pi^2}  \lesssim 1 \rightarrow |\alpha_m|^{EFT} \lesssim 2\times 10^{-2} \ .
\label{alim}
\eea
 Note that for values of $\alpha_m$ close to this bound, and negative, cancellations could lead to a value of  the field below the Planck mass,
 \bea
 \phi < \bar M_P \textrm{ for } N \simeq 60 \ ,
 \eea 
 while there would be no simultaneous cancellation in the maximum value of the potential
 \bea
 V_N\simeq \frac{\tilde m^2 N}{2 \pi} \left( 1+ \frac{5}{3} \, \frac{N^2\alpha_m}{4 \pi^2} \right) \ .
 \eea
 Also note that the condition in Eq.~\ref{alim} limits shift in the slow-roll parameters Eqs.~\ref{eps} and \ref{eta} to be less than $(3-8)\times10^{-2}$, respectively.
 
  \begin{figure}[t!]
\begin{minipage}{8in}
\hspace*{-0.7in}
\centerline{\includegraphics[height=9cm]{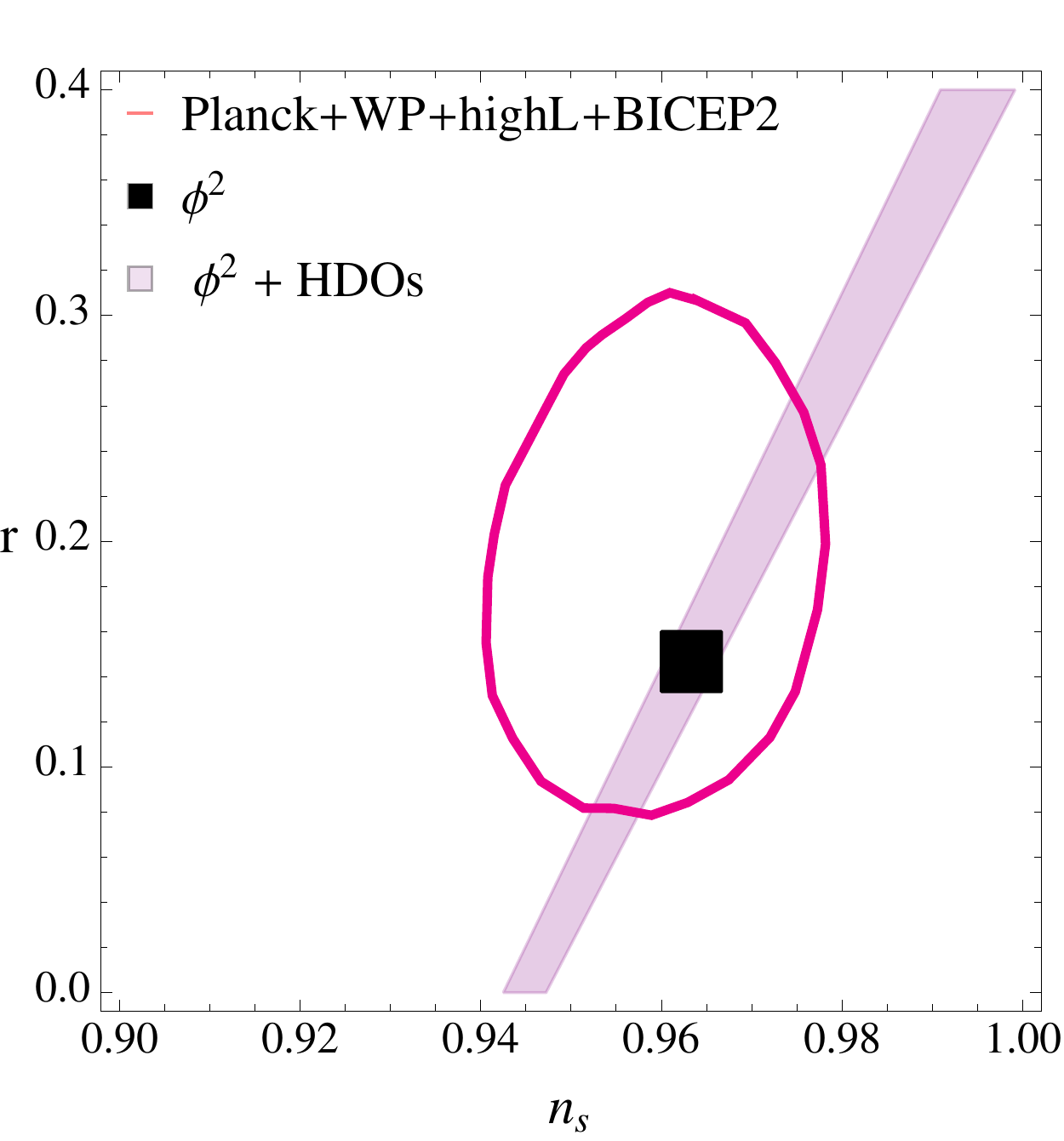}} \label{figc}
\hfill
\end{minipage}
\caption{
{\it The effect of higher-dimensional operators is shown by the purple band, the black box corresponds to chaotic inflation without higher-dimensional operators, and the pink circle is the area of 95\% CL from BICEP2.
} 
}
\end{figure}

 The higher-dimensional operators would also affect the scalar power spectrum
 \bea
 P_{\cal R}^{1/2} = \frac{4 \sqrt{24 \pi} }{3} \frac{\, V(\tilde\phi_N)^{3/2}}{V'(\tilde\phi_N)} \simeq P_{\cal R,CI}^{1/2}  \,\left( 1- \frac{5}{6} \frac{N^2 \alpha_m}{4 \pi^2}\, \right) \ ,
 \eea
 where $P_{\cal R,CI}^{1/2}=2 \sqrt{\frac{2}{3 \pi}} N \tilde m $ is the value obtained in chaotic inflation with no higher-dimensional operators. Hence the same limit for $\tilde m$ is obtained in this case,
 \bea
 \tilde m \simeq 4 \times 10^{-7} \rightarrow m \sim 10^{12} \textrm{ GeV.}  
 \label{mval}
 \eea
 
 The value of the spectral index in the presence of the dimension-six operator is
 \bea
 n_s-1= (n_s-1)_{CI} \, \left( 1-\frac{5}{3} \, \frac{N^2 \alpha_m}{4 \pi^2}\right) \ .
 \eea 
 Inspecting Eq.~\ref{alim}, one obtains that the shift on $n_s-1$ is bounded by
 \bea
 \Delta(n_s-1) \lesssim \frac{5}{3} \ .
 \eea
 Similarly for the tensor-to-scalar ratio one obtains
 \bea
 r=r_{CI}  \, \left( 1+\frac{10}{3} \, \frac{N^2 \alpha_m}{4 \pi^2}\right) \ .
 \eea
 The shift in values of   $n_s-1$ and $r$ is constrained by the latest BICEP2 combination, as shown in Fig.~\ref{figc}. The purple band shows the effect of non-zero higher-dimensional operators, the black box corresponds to chaotic inflation without higher-dimensional operators, and the pink circle is the area of 95\% CL from BICEP2. Note we have neglected possible effects on the running of Planck results on the combination, but we expect these to be negligible.

The region of the parameter space allowed by BICEP2's combination is
\bea
\alpha_m^{BICEP2} \in [-2,3] \times 10^{-3} \ .
\label{alimf}
\eea
Comparing with the condition for a valid expansion, Eq.~\ref{alim}, one can see that the data already constrains the higher-dimensional operators one order of magnitude below the limit of validity of the effective theory. 

The limit on the coefficient of the higher-dimensional operator operator $c_6$ can be read by Eqs.~\ref{aldef} and \ref{mval},
\bea
c_6  \lesssim 10^{-9} \ .
\label{lc6}
\eea

Let us finish this section by discussing the effect of higher-dimensional operators in the scale of inflation. The value of $H_I$ is given by
\bea
H_I \simeq \pi \bar M_P \, \sqrt{\frac{2.2 \times 10^{-19} \, r_{CI}}{2}} \, \left( 1+\frac{5}{3} \, \frac{N^2 \alpha_m}{4 \pi^2}\right) \ ,
\eea
which would, in principle, allow a cancellation between the two terms in parenthesis, and potentially a lower value for the scale of inflation. Nevertheless, the constrain in Eq.~\ref{alimf} excludes this possibility.

\section{The effect of higher-dimensional operators in $\lambda \phi^4$}

 \begin{figure}[t!]
\begin{minipage}{8in}
\hspace*{-0.7in}
\centerline{\includegraphics[height=9cm]{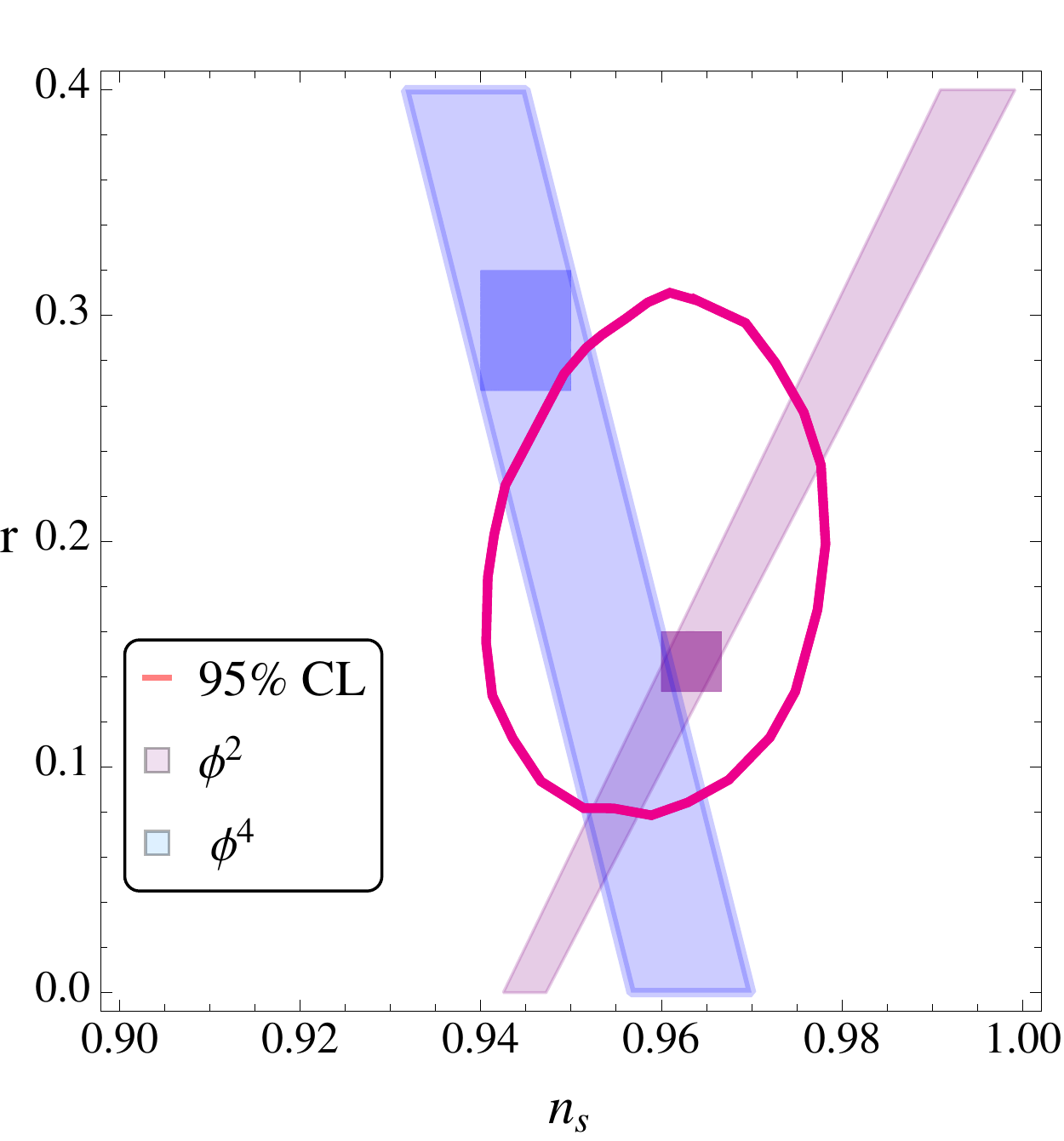}} \label{figc42}
\hfill
\end{minipage}
\caption{
{\it The effect of higher-dimensional operators on $\phi^4$ and $\phi^2$ potentials shown in blue and purple respectively. The darkerboxes corresponds to potentials without higher-dimensional operators, and the pink circle is the area of 95\% CL from BICEP2.
} 
}
\end{figure}

In this section we will follow the same steps as for chaotic inflation. Starting with the potential
\bea
V(\tilde\phi) = \bar M_P^{4} \left( \lambda \tilde \phi^4 + c_n \tilde \phi^n \right).
\eea
We consider dimension-six operators as the first correction to the normalizable potential, and define a parameter $\alpha_\lambda$ as
\bea
c_6 = \alpha_\lambda \lambda \rightarrow V(\tilde\phi) = \bar M_P^{4} \lambda \tilde \phi^4  \left( 1+ \alpha_\lambda \tilde \phi^{2} \right) \ ,
\label{aldef2}
\eea
hence, the validity of the effective theory requires
\bea
 |\alpha_\lambda|\tilde \phi^{2} < 1 \ .
 \label{condexp2}
\eea

The slow-roll parameters adopt the following form
\bea
\epsilon &=& \epsilon_{\phi^4} \, (1+\alpha_\lambda \tilde\phi^2) \\
\eta &=& \eta_{\phi^4} \, (1+5 \alpha_\lambda \tilde\phi^2/2) 
\eea
where $\epsilon_{\phi^4}=1/(\pi \tilde\phi^2)$ and $ \eta_{\phi^4} =1/(2 \pi \tilde \phi^2)$. The value of $\tilde\phi_N$ is
\bea
\tilde\phi_N^2 = \frac{N}{\pi} \, \left( 1+ \frac{N \alpha_\lambda}{4 \pi}\right) \ , 
\eea
which in Eq.~\ref{condexp2} implies
 \bea
 |\alpha_\lambda|^{EFT} \lesssim 0.06
\eea
In Fig.~\ref{figc42} we show the comparison between the regions for $\phi^4$ and $\phi^2$ potentials shown in blue and purple respectively. The darkerboxes corresponds to potentials without higher-dimensional operators, and the pink circle is the area of 95\% CL from BICEP2. This translates into a limit on $\alpha_\lambda$ close to the EFT limit,
\bea
\alpha^{BICEP}_\lambda  \in [-0.06,0] \to c_6 < 10^{-15}
\eea
and a limit on the dimension-six operator which is more stringent than in the case of chaotic inflation, Eq.~\ref{lc6}.

\section{ Conclusions}

Our results have important implications for models of inflation. We have shown that quantum gravitational effects parametrized by higher dimensional operators have a significant impact in the interpretation of the data coming from observations of the PLANCK satellite or the BICEP2 experiment, even in a region of parameter space well within the validity of the effective theory approach. 

One possible interpretation is that quantum gravity effects make it very difficult to probe a specific model of inflation. This observation is similar to that made in the case of grand unified theories \cite{Calmet:2008df}. Another interpretation is that quantum gravitational effects can salvage models which seemed to be disfavoured by current data such as e.g. Higgs inflation \cite{Bezrukov:2007ep,Calmet:2013hia,Atkins:2010yg}. Within a specific models of inflation we can, as we have shown, derive bounds on higher dimensional operators and more specifically on the Wilson coefficients of these operators. In that sense the cosmic microwave background provides an ideal environment to probe quantum gravity effects. Indeed, we can now hope to probe the symmetries of quantum gravity. For example, we could answer questions such as: is there an approximate shift symmetry which prevents these higher dimensional operators? Are Lorentz invariance \cite{Colladay:1998fq} and CPT  invariance \cite{Colladay:1996iz} valid symmetries at the Planck scale? Is space-time quantized and is there a minimal length in nature as expect from a unification of quantum mechanics and general relativity \cite{Calmet:2007vb} or is general relativity a purely classical theory \cite{Ashoorioon:2012kh}?

Our results show that we will never be able to probe the potential of the inflaton without making assumptions about quantum gravity, or rather without assuming a specific framework for quantum gravity. Nevertheless, our results indicate that an effective theory approach is a valuable tool thanks to the outstanding degree of precision achieved by the experiments. 

Note that we focussed here on higher dimensional operators suppressed by the Planck scale, but if the inflaton is embedded into a grand unified theory, there is another natural scale below the Planck scale which would force any Wilson coefficient to be even tinier than the ones we discussed, namely the unification scale which at $10^{16}$ GeV coincides with the scale of inflation as mentioned earlier~\cite{Dimopoulos:1997fv}. Finally, this work can easily be extended to other models of inflations such as hybrid inflation and other higher dimensional operators. 

{\it Acknowledgments:}
This work is supported in part by the European Cooperation in Science and Technology (COST) action MP0905 ``Black Holes in a Violent  Universe" and by the Science and Technology Facilities Council (grant number  ST/J000477/1).


\bigskip{}

\end{document}